\documentclass[a4paper,useAMS,usenatbib,letters]{mn2e}

\usepackage{graphicx}  
\usepackage{dcolumn}   
\usepackage{bm}        
\usepackage{amsfonts,amsmath,amssymb,mathrsfs}
\usepackage{color}

\newcommand{\bea}{\begin{eqnarray}}
\newcommand{\eea}{\end{eqnarray}}
\newcommand{\beq}{\begin{equation}}
\newcommand{\eeq}{\end{equation}}

\newcommand{\cf}{\textit{cf.}}
\newcommand{\ie}{\textit{i.e.}~}
\newcommand{\eg}{\textit{e.g.}~}
\newcommand{\ms}{{\rm ms}}
\newcommand{\km}{{\rm km}}

\newcommand{\khz}{{\rm kHz}}

\newcommand{\G}{{\rm G}}

\begin{document}

\title[Detecting magnetic fields during BNS inspiral]{Can magnetic fields
  be detected during the inspiral of binary neutron stars?}

\author[Giacomazzo, Rezzolla \& Baiotti]{Bruno Giacomazzo,$^1$
  Luciano Rezzolla$^{1,2}$ and Luca
  Baiotti$^3$\\
$^1$Max-Planck-Institut f\"ur Gravitationsphysik,
    Albert-Einstein-Institut, Potsdam-Golm, 14476, Germany\\
$^2$Department of Physics and
    Astronomy, Louisiana State University, Baton Rouge, LA 70803, USA\\
$^3$Graduate School of Arts and Sciences, University
    of Tokyo, Komaba, Meguro-ku, Tokyo, 153-8902, Japan}

\maketitle

\begin{abstract}
  Using accurate and fully general-relativistic simulations we assess
  the effect that magnetic fields have on the gravitational-wave
  emission produced during the inspiral and merger of magnetized
  neutron stars. In particular, we show that magnetic fields have an
  impact after the merger, because they are amplified by a Kelvin-Helmholtz
  instability, but \textit{also} during the inspiral, most likely because the
  magnetic tension reduces the stellar tidal deformation for extremely
  large initial magnetic fields, $B_{0}\gtrsim 10^{17}\,{\rm G}$. We
  quantify the influence of magnetic fields by computing the overlap,
  ${\cal O}$, between the waveforms produced during the inspiral by
  magnetized and unmagnetized binaries. We find that for any realistic
  magnetic field strength $B_{0}\lesssim 10^{14}\,{\rm G}$ the overlap
  during the inspiral is ${\cal O}\gtrsim 0.999$ and is quite
  insensitive to the mass of the neutron stars. Only for
  unrealistically large magnetic fields like $B_{\rm 0}\simeq 10^{17}\,{\rm
    G}$ the overlap does decrease noticeably, becoming at our
  resolutions ${\cal
    O}\lesssim 0.76/0.67$ for stars with baryon masses $M_{b} \simeq
  1.4/1.6\,M_{\odot}$, respectively. Because neutron stars are
  expected to merge with magnetic fields $\sim
  10^{8}-10^{10}\,{\rm G}$ and because present detectors are
  sensitive to ${\cal O}\lesssim 0.995$, we conclude that it is very
  unlikely that the present
  detectors will be able to discern the presence of magnetic fields
  during the inspiral of neutron stars.
\end{abstract}

\begin{keywords}
  relativity -- gravitational waves -- stars: neutron -- binaries:
  general -- magnetic fields -- MHD
\end{keywords}

\section{Introduction}

Numerous astronomical observations suggest that large magnetic fields
are associated with neutron stars (NSs). Indeed, evidence for the
existence of binary NSs is obtained from binary pulsars, in
which one or both NSs are seen to have a large magnetic
field. In General Relativity such binary systems cannot be stationary 
because they emit gravitational waves (GWs) which extract
energy and angular momentum from the binary, inducing it to inspiral
and merge. During the final stages of the inspiral the
GW emission is expected to be strong enough to be
relevant for the detectors now operative at design sensitivities and
it promises to provide important information on the equation of state
(EOS) regulating the NS matter~\citep{Read2009}. In addition
to their importance as sources of GWs, however, the
merger of binary NSs is likely to provide important
information on the physics of short gamma-ray bursts (GRBs). The
coalescence of the two NSs, in fact, gives rise, either
promptly or after some interval, to a system composed of a torus
orbiting around a rapidly rotating black
hole (BH)~\citep{Baiotti08,Yamamoto2008}. The complex plasma physics
accompanying this event is probably behind the ``engine'' powering
GRBs~\citep{Piran:2004ba,Meszaros:2006rc}.

There is little doubt, therefore, about the importance of assessing
the role played by magnetic fields in the inspiral and merger of
binary NSs. Yet, determining this accurately is a remarkably difficult
task requiring the solution of the Einstein equations together with
those of general-relativistic magnetohydrodynamics (GRMHD). So far,
only two GRMHD simulations have been
reported~\citep{Anderson2008,Etienne08}, reaching different
conclusions about the importance of very strong magnetic fields
($B\sim 10^{16}-10^{17}\,\G$). The aim of this Letter is to go beyond
these qualitative estimates and provide a first quantitative
measurement of the influence of magnetic fields on both the inspiral
and the merger of magnetized NSs.

By considering a large range of magnetic fields, which includes
values more realistic than those used in the works cited above,
and two different masses, we find that magnetic fields
generally grow \textit{after} the merger, when the turbulent motions
triggered during the merger by the Kelvin-Helmholtz (KH) instability
amplify any initial poloidal magnetic field producing a toroidal one
whose strength rapidly becomes comparable to the poloidal one. In addition, we
find that magnetic fields can, at least in principle, play a role
already \textit{during the inspiral} if sufficiently strong. This is
most likely due to the magnetic tension, which decreases the NS deformability,
increases the compactness, and thus delays the time of merger. In
practice, however, the influence of magnetic fields during the
inspiral appears only for values $\sim 10^{17}\,\G$ which are
unrealistic~\citep{UGK98,Abdolrahimi2009}. As a result, it is very
unlikely that present detectors will be able to measure the presence
of magnetic fields during the inspiral of NSs.
Finally, we show that high-order numerical schemes are essential to
draw robust conclusions while lower-order
methods incorrectly suggest that even strong magnetic fields have no
influence at all.

\section{Mathematical and numerical setup}

All the results presented here were computed by solving the GRMHD
equations in the ideal MHD approximation (\ie assuming an infinite
electrical conductivity) and in dynamical spacetimes. The evolution 
of the spacetime was obtained using the
\texttt{CCATIE} code, a three-dimensional finite-differencing code
providing a solution of a conformally traceless formulation of the
Einstein equations~\citep{Koppitz-etal-2007aa,Pollney:2007ss}. The
GRMHD equations were instead solved using the \texttt{Whisky} code
presented in~\citet{Giacomazzo:2007ti}, thus adopting a
flux-conservative formulation of the GRMHD equations~\citep{Anton05}
and high-resolution shock-capturing schemes. In particular, we have
computed the fluxes using the Harten-Lax-van Leer-Einfeldt (HLLE)
approximate Riemann solver~\citep{Harten83}, while the reconstruction
was made using the $3$rd-order piecewise parabolic method~\citep{Colella84}. Furthermore, to guarantee
the divergence-free character of the MHD equations we have employed
the flux-constrained-transport approach~\citep{Toth2000}. The code has
been validated against a series of tests in special
relativity~\citep{Giacomazzo:2005jy} and in full general relativity
[see~\citet{Giacomazzo:2007ti}].

The system of GRMHD equations is closed by an EOS and, as discussed in
detail in~\citet{Baiotti08}, the choice of the EOS plays a fundamental
role in the post-merger dynamics and significantly influences the
survival time against gravitational collapse of the hyper-massive
neutron star (HMNS) likely produced by the merger. It is hence
important that special attention is paid to use EOSs that are
physically realistic, as done in~\citet{Oechslin07b} within a
conformally-flat description of the fields and a simplified treatment of
the hydrodynamics. Because we are here mostly concerned with computing
a first quantitative estimate of the role played by magnetic fields
rather than with a realistic description of the NS matter,
we have employed the commonly used ``ideal-fluid'' EOS in which the
pressure $P$ is expressed as $P = \rho\, \epsilon(\Gamma-1) $, where
$\rho$ is the rest-mass density, $\epsilon$ is the specific internal
energy and $\Gamma$ is the adiabatic exponent. While simple, such EOS
provides a reasonable approximation and we expect that the use of
realistic EOSs would not change the main results of this work.

All equations are solved on a Cartesian grid using the
vertex-centered mesh-refinement scheme provided by the \texttt{Carpet}
driver~\citep{Schnetter-etal-03b}. Differently from~\citet{Baiotti08}, 
we use here larger fixed refined grids rather than smaller moving ones. While computationally more
expensive, this choice reduces the violations in the divergence of the
magnetic field due to interpolations in the buffer zones between
refinement levels. In this way, the divergence of the magnetic field
on the finest grid (not including the buffer zones) is zero to machine
precision.

We have used five refinement levels with a 180-degree rotational
symmetry around the $z$ axis and a reflection symmetry across the
$z=0$ plane (in practice, we simulate only the region $\{x\geq0,z\geq0\}$).
The finest grid has a resolution of $h=354.4 \,{\rm m}$ and extends up
to $r=44 \,\km$; the coarsest grid has $h=5.6704\,\km$ and extends up
to $r=380 \,\km$. Our finest grid therefore contains both NSs at all
times and each NS is covered with $\approx 80^3$ points.
In~\citet{Anderson2008} the finest grid had $h=0.46\,\km$, thus with
$\approx 70^3$ points across each star, and~\citet{Etienne08} had
$\gtrsim 40^3$ points. So our resolution is higher than that in the
above works, but it is only half of that in~\citet{Baiotti08}, and it
is barely sufficient to reach convergent results for the inspiral.

\section{Initial data.}

The initial data are the same as used in~\citet{Baiotti08} and were
produced by~\citet{Taniguchi02b} with the multi-domain spectral-method
code {\tt LORENE} (\texttt{http://www.lorene.obspm.fr}). The
initial solutions for the binaries are obtained assuming a
quasi-circular orbit, an irrotational velocity field, and a
conformally-flat spatial metric. The matter is modelled using a
polytropic EOS $P = K \rho^{\Gamma}$ with $K=123.6$ and
$\Gamma=2$. Since no self-consistent solution is available yet for
magnetized binaries, a poloidal magnetic field is added a-posteriori
using the vector potential $A_{\phi} \equiv \varpi^2 A_b\, {\rm max}\,(P-P_{\rm cut},0)^{n_{\rm s}}$, where $\varpi \equiv \sqrt{x^2+y^2}$, $A_b>0$ parameterizes the
strength of the magnetic field, $P_{\rm cut}$ defines where in the
NS the magnetic field goes to zero, and $n_{\rm s}$
determines the smoothness of the potential. The components of the
magnetic field are then computed by taking the curl of the Cartesian
components of~$A_{\phi}$ to enforce that the divergence of the
magnetic field is zero to machine precision. Here we set $P_{\rm
  cut}=0.04\, {\rm max}(P)$, and $n_{\rm s}=2$ to enforce that both
the magnetic field and its first derivative are zero at $P=P_{\rm
  cut}$. In~\citet{Anderson2008} the magnetic field was built with an
equivalent expression but with $P_{\rm cut}$ set to the pressure in
the atmosphere, and in~\citet{Etienne08} the expression used is only
slightly different, but $P_{\rm cut}$ is set to be $4\%-0.1\%$ of
${\rm max}(P)$. Both above works set
$n_{\rm s}=1$. Note that the magnetic fields are confined at all times
inside the NS matter and hence they cannot ``repel'' each
other during the inspiral, as claimed in~\citet{Anderson2008}.

\begin{table*}
\begin{minipage}{180mm}
\caption{\label{tab:id}Properties of the eight equal-mass binaries considered.}
\begin{tabular}{lccccccccccc}
Binary &
\multicolumn{1}{c}{$B_{0}$} &
\multicolumn{1}{c}{$d/M_{_{\rm ADM}}$} &
\multicolumn{1}{c}{$M_{b}$} &
\multicolumn{1}{c}{$M_{_{\rm ADM}}$} &
\multicolumn{1}{c}{$J$} &
\multicolumn{1}{c}{$\Omega_0$} &
\multicolumn{1}{c}{$r_e$} &
\multicolumn{1}{c}{$r_p/r_e$}&
\multicolumn{1}{c}{$\rho_{\rm max}$}& 
\multicolumn{1}{c}{$(M_{*}/R)_{\infty}$}
\\
&
\multicolumn{1}{c}{$({\rm G})$} 
&
&
\multicolumn{1}{c}{$(M_{\odot})$} &
\multicolumn{1}{c}{$(M_{\odot})$} &
\multicolumn{1}{c}{$({\rm g\, cm^2\, s^{-1}})$} &
\multicolumn{1}{c}{$({\rm rad\, ms^{-1}})$} &
\multicolumn{1}{c}{$({\rm km})$} &
\multicolumn{1}{c}{}&
\multicolumn{1}{c}{$({\rm g\, cm^{-3}})$}& 
\multicolumn{1}{c}{}
\\
\hline
${\rm M}1.45$-B$*$  & $0$ or $1.97 \times10^{*}$ & $14.3$ & $1.445$ & $2.681$ & $6.5083\times10^{49}$ & $1.78$ & $15.2$ & $0.899$ & $4.58\times10^{14}$ & 0.12 \\
${\rm M}1.62$-B$*$  & $0$ or $1.97 \times10^{*}$ & $13.2$ & $1.625$ & $2.982$ & $7.7805\times10^{49}$ & $1.85$ & $13.7$ & $0.931$ & $5.91\times10^{14}$ & 0.14 \\
\end{tabular}

\medskip

{\it Notes.} The different columns refer, respectively, to model
name; maximum initial magnetic field $B_{0}$, where $*$ is either $0$
(in which case $B_0=0$), $12, 14$ or $17$; proper separation between the stellar centres $d/M_{_{\rm
    ADM}}$; baryon mass $M_{b}$ of each star; total ADM mass $M_{_{\rm
    ADM}}$; angular momentum $J$; initial orbital angular velocity
$\Omega_0$; mean coordinate radius $r_e$ along the line connecting
the two stars; ratio of the polar to the equatorial coordinate
radii $r_p/r_e$; maximum rest-mass density $\rho_{\rm max}$; compactness of the stars $(M_{*}/R)_{\infty}$.
\end{minipage}
\end{table*}

Table~\ref{tab:id} lists some of the properties of the eight
equal-mass binaries studied here. In more detail, we 
consider two classes of binaries differing in the initial masses,
\ie binaries ${\rm M}1.45$-${\rm B}*$ (or low-mass), and binaries
${\rm M}1.62$-${\rm B}*$ (or high-mass). For each of these classes we
take four different magnetizations (indicated by the
asterisk) so that, for instance, ${\rm M}1.45$-${\rm B12}$ is a
low-mass binary with a maximum initial magnetic field $B_{\rm 0}=1.97
\times 10^{12}\,\G$ [the binaries with zero magnetic fields are the
  same as those evolved in~\citet{Baiotti08}]. In summary,
we consider 8 different binaries that are either unmagnetized or with
magnetic fields as large as $B_{0} \simeq 10^{17}\,\G$. We
reinforce a remark already made in the Introduction: all astronomical
observations and theoretical considerations suggest that the magnetic
fields in binary NS systems just before the merger are
much smaller than $\simeq 10^{17}\,\G$. However, we do consider such
large fields here, firstly to compare with those studied 
by~\citet{Anderson2008} and by~\citet{Etienne08},
and secondly because by doing so we can determine important lower
limits on the detectability of magnetic fields during the inspiral.

\section{Gravitational waves and overlaps}

\begin{figure}
  \begin{center}
  \includegraphics[angle=0,width=0.37\textwidth]{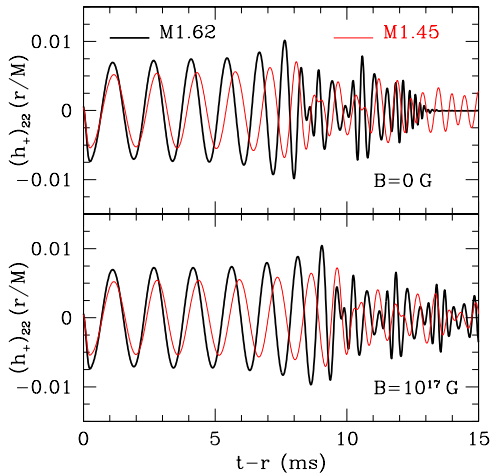}
  \end{center}
  \caption{\label{fig1}\textit{Top panel:~}$\ell=2,\,m=2$ component of
    the $h_+$ polarization from binaries with different masses
    (thick line: $1.62\,M_{\odot}$; thin line: $1.45\,M_{\odot}$)
    and zero magnetic field. \textit{Bottom panel:~}the same but for 
    binaries with an initial magnetic field $B_0
    \simeq 10^{17}\,\G$.}
\end{figure}

We postpone the discussion on the matter dynamics to a subsequent
paper and concentrate here on the GW emission. A
representative summary is offered in Fig.~\ref{fig1}, which reports
the $\ell=2,\,m=2$ component of the $h_+$ polarization (modulo a phase
difference, $h_{\times}$ shows the same behaviour). More specifically,
Fig.~\ref{fig1} highlights the differences in the GWs
from binaries with different masses, namely $1.62\,M_{\odot}$ (thick
 line) and $1.45\,M_{\odot}$ (thin line), when the initial magnetic
field is either zero (top panel) or as high as $\simeq
10^{17}\,\G$ (bottom panel). Figure~\ref{fig1} clearly shows that when
the NSs are not magnetized, the high-mass binary has a
larger-amplitude GW emission, it experiences an
earlier merger and the HMNS at these resolutions collapses to a
rapidly rotating BH after only $\sim\!4\,\ms$, while the HMNS
from the low-mass binary does not collapse. In contrast, when the
NSs are initially magnetized, the strong magnetic tension
most likely reduces the tidal deformations and results in a delayed merger
time (defined as the time when the maximum
  rest-mass density has a first significant minimum; \cf~Fig~2 or 8
  in~\citet{Baiotti08}). Furthermore, the additional pressure support
coming from the intense magnetic fields is such that neither the
high-mass nor the low-mass binary collapse promptly to a BH
over the $\sim 15\,\ms$ of the simulations (see the bottom panel of
Fig.~\ref{fig4} for a comparison of GWs for high-mass
binaries with different magnetic fields). Overall, Fig.~\ref{fig1}
shows that magnetic fields have a strong impact on the
GWs emitted after the merger but \textit{also} during the inspiral,
if sufficiently strong. If, on the other hand, the magnetic fields are
more realistic, \eg $\sim 10^{12}\,\G$, then differences appear
\textit{only} after the merger.
For compactness, comparisons of this type cannot be presented here but
will appear in a longer companion paper (Baiotti et al., in preparation).

\begin{figure}
  \begin{center}
  \includegraphics[angle=0,width=0.37\textwidth]{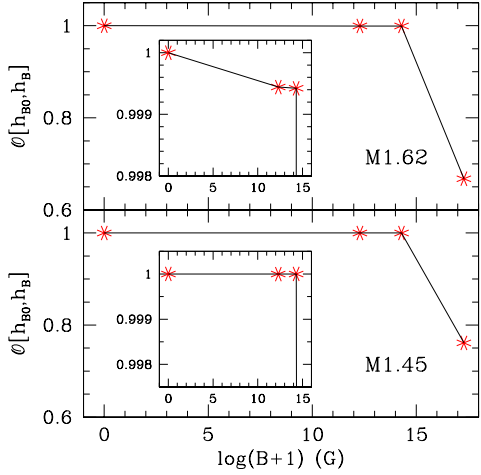}
  \end{center}
  \caption{\label{fig2}Overlap in inspiral waveforms from binaries
    with different magnetization; the top and
    bottom panels refer to the high- and low-mass binaries,
    respectively.}
\end{figure}

While generic, this behaviour depends sensitively on the strength of
the initial magnetic field and there exists a critical magnetic field
below which the MHD effects during the inspiral are not important. In
order to quantify this we have computed the overlap between two
waveforms $h_{_{\rm B1}},~h_{_{\rm B2}}$ from binaries with initial
magnetic fields ${\rm B1},~{\rm B2}$ as
\begin{equation}
\label{overlap}
\mathcal{O}[h_{_{\rm B1}}, h_{_{\rm B2}}] \equiv \frac{\langle h_{_{\rm
      B1}} | h_{_{\rm B2}} \rangle}{\sqrt{\langle h_{_{\rm B1}} |
    h_{_{\rm B1}} \rangle \langle h_{_{\rm B2}} | h_{_{\rm B2}}
    \rangle}}\,,
\end{equation}
where $\langle h_{_{\rm B1}} | h_{_{\rm B2}} \rangle$ is the scalar
product and is defined as
\begin{equation}
\label{scalarproduct}
\langle h_{_{\rm B1}} | h_{_{\rm B2}} \rangle \equiv 4 \Re
	\int_0^\infty df \frac{\tilde{h}_{_{\rm B1}}(f) \tilde{h}_{_{\rm
	B2}}^*(f)}{S_h(f)}\,,
\end{equation}
and $\tilde{h}(f)$ is the Fourier transform of the GW
$h(t)$ and $S_h(f)$ is the noise power spectral density of the
detector (we have here considered LIGO). Clearly, waveforms that are
very similar have ${\cal O}\simeq 1$. A general view is shown in
Fig.~\ref{fig2}, which reports the overlaps between the unmagnetized
binaries and binaries with different magnetizations, \ie
$\mathcal{O}[h_{_{\rm B0}}, h_{_{\rm B}}]$, for the two masses
considered here (top and bottom panels, respectively). Note that the
overlap is relative to the inspiral only [\ie the
  integral~\eqref{scalarproduct} is cut off at the orbital frequency
  at merger] since this is the phase for which our results are
convergent (becoming only consistent after the merger as a result of
the development of turbulence) and because the post-merger evolution
can only further decrease ${\cal O}$. It is evident that for the
high-mass binary (top panel) the influence of the magnetic field is
noticeable only for very large magnetic fields (${\cal O}\simeq
0.999$ for $B_{0}\simeq 10^{14}\,\G$ and ${\cal
  O}\simeq 0.668$ for $B_{0}\simeq 10^{17}\,\G$). This is true also for
the low-mass binary (bottom panel) whose smaller compactness, however,
leads to larger overlaps (\ie ${\cal O}\simeq 0.761$ for $B_{0}\simeq
10^{17}\,\G$). In view of these results, of the fact that
NSs just prior to merger are expected to have magnetic fields
$\sim
10^8-10^{10}\,\G$~\citep{UGK98,Abdolrahimi2009}, and that present
detectors are sensitive to ${\cal O}\lesssim
0.995$~\citep{Lindblom08}, we conclude that it is very unlikely that
the presence of realistic magnetic fields can be detected during the
inspiral.

\section{Magnetic field amplification}

\begin{figure}
  \begin{center}
  \includegraphics[angle=0,width=0.37\textwidth]{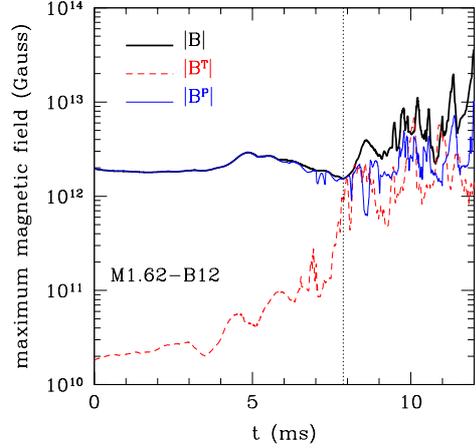}
  \end{center}
  \caption{\label{fig3}Evolution of the maxima of the magnetic field $|B|\equiv
    (B^iB_i)^{1/2}$ (thick solid line) and of its toroidal $|B^{\rm T}|$
    (dashed line) and poloidal $|B^{\rm P}|$ (thin solid line)
    components for the high-mass case with $B_{\rm 0}\simeq
    10^{12}\,\G$. The dotted line marks the merger time.}
\end{figure}

As discussed in detail in~\citet{Baiotti08}, during the merger a shear
layer develops in the region where the two NSs enter in
contact. Across this layer, the tangential components of the velocity
are discontinuous and this leads to the development of a KH
instability and thus to the production of vortices [\cf~Fig. 16
  of~\citet{Baiotti08}]. When poloidal magnetic fields are present,
this hydrodynamical instability can lead to exponentially growing
toroidal magnetic fields, thus increasing the energy stored in
magnetic fields. This mechanism, already observed in Newtonian
simulations~\citep{Price06} but not before in general relativistic
ones, is likely to be important for explaining the physics powering
short GRBs. Taking ${\rm M}1.62$-${\rm B}12$ as a reference,
Fig.~\ref{fig3} shows the evolution of the maxima of the magnetic
field $|B|\equiv (B^iB_i)^{1/2}$ (thick line), and of its toroidal
$|B^{\rm T}|$ (dashed line), and poloidal $|B^{\rm P}|$ (thin
line) components. A vertical dotted line marks the merger, occurring
$\approx 1\,\ms$ after the KH instability has started
developing. Clearly, as long as the KH instability is active, the
toroidal magnetic field is amplified exponentially, until it reaches
values comparable to the poloidal one (this is different and more
reasonable than what found by~\citet{Price06}, where the magnetic
field reached energy equipartition values). Note that the magnetic
field grows considerably also when the HMNS collapses to a BH as a
result of magnetic-flux conservation in the collapsing NS matter.

As discussed in~\citet{Price06} and in~\citet{Baiotti08}, much in the
development of the KH instability and in the subsequent magnetic field
amplification depends on the resolution used. A detailed study of the
turbulent regime and magnetic field amplification produced by the
merger is extremely challenging and requires resolutions well above
the ones that can be afforded now in GRMHD simulations. Nevertheless,
on the basis of preliminary investigations with different resolutions, we expect the behaviour in Fig.~\ref{fig3} to be qualitatively correct
and, hence, that as long as the KH is active, the poloidal magnetic
field is coiled into a toroidal one, increasing it exponentially to
values comparable with the poloidal one. When equipartition among the
components is reached, the large magnetic tension suppresses the KH
instability, preventing a further growth of the toroidal magnetic
field. An analysis of this process will be presented elsewhere.

\section{The importance of high-order methods}

\begin{figure}
  \begin{center}
  \includegraphics[angle=0,width=0.37\textwidth]{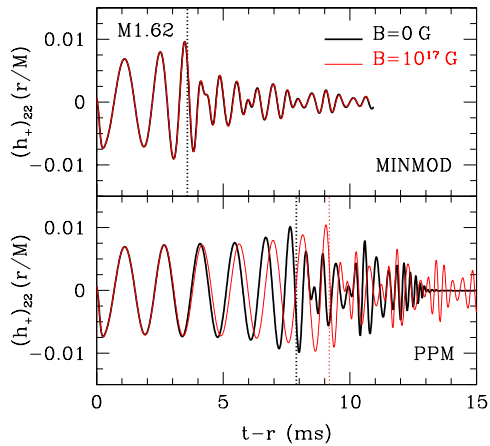}
  \end{center}
  \caption{\label{fig4}GW emission from the the
    high-mass binary evolved using either a $2$nd-order scheme (top
    panel) or a $3$rd-order one (bottom panel). A thick (thin) line
    refers to a binary with $B_{0}=0\,\G$ ($10^{17}\,\G$), while the
    vertical lines indicate the time of merger.}
\end{figure}

As shown in~\citet{Baiotti08}, the use of reconstruction schemes of
sufficiently high order is \textit{essential} for a correct
calculation of the GW signal. This is stressed also in
Fig.~\ref{fig4}, which presents a comparison in the GW emission from
the high-mass binary for evolutions made using either a
$2$nd-order MINMOD scheme (top panel) or a $3$rd-order PPM one (bottom
panel) with the same grid structure and resolution used in the previous runs. In both cases a thick line refers to the unmagnetized binary
while a thin line to the binary with $B_{0}\simeq
10^{17}\,\G$. Clearly, while the evolutions with MINMOD show only
minimal differences between the magnetized and unmagnetized case
(${\cal O}=0.9994$ over the whole waveform), the
evolutions using PPM show considerable differences (${\cal
  O}=0.6500$), both during the inspiral and after the
merger. More precisely, although we use exactly the same initial data,
the binaries evolved with MINMOD merge almost two orbits earlier than
those evolved with PPM [\cf~vertical dotted lines in
  Fig.~\ref{fig4}]. Additionally, the unmagnetized binary evolved
with MINMOD does not collapse to a BH, in contrast to what happens
when using PPM at these resolutions. These differences are due to the
numerical dissipation of the $2$nd-order method, which is inadequate
at these resolutions. This could explain why the calculations
in~\citet{Etienne08}, where a $2$nd-order reconstruction and a lower
resolution were used, show only small differences between unmagnetized
and magnetized binaries.

\section{Conclusions}

We have presented accurate simulations of the inspiral and merger of
magnetized NSs and found that magnetic fields have an impact
\textit{both} during the inspiral and after the merger but only if
sufficiently strong. Comparing waveforms for different magnetizations
we have found that for $B_{0}\lesssim 10^{14}\,{\rm G}$, the overlap
${\cal O}\gtrsim 0.999$ and is quite insensitive to the mass of the
NSs. Only for unrealistically large magnetic fields ($B_{\rm
  0}\simeq 10^{17}\,{\rm G}$), the overlap decreases noticeably,
becoming at our resolutions ${\cal O}\lesssim 0.76/0.67$ for stars with baryon masses
$M_b \simeq 1.4/1.6\,M_{\odot}$, respectively. Since the magnetic
fields in NSs just prior to merger are expected to be rather
small ($\sim 10^8-10^{10}\,\G$)~\citep{UGK98,Abdolrahimi2009}, we
conclude that it is very unlikely that the present
detectors will be able to measure the presence of
magnetic fields during the inspiral. Magnetic fields could be however
detectable after the merger and hence in the part of the spectrum at
frequencies $\gtrsim 2\khz$ [\cf~spectra in~\citet{Baiotti08}].

Another important result discussed here is the evidence that a KH
instability develops during the merger, leading to the exponential growth of
a toroidal magnetic field, the strength of which becomes comparable with the
poloidal one. This additional magnetic field can modify
the structure of the HMNS and decrease the overlap after the
merger. Finally, we have provided concrete evidence that high-order
methods are \textit{essential} to draw robust
conclusions and that instead lower-order methods incorrectly suggest
that magnetic fields have no influence at all.

\section*{Acknowledgments}
We thank the developers of \texttt{Lorene} for the initial data and
those of \texttt{Carpet} for the mesh refinement. Special thanks go to
our late colleague and friend Thomas Radke. Useful input from
C.~Palenzuela, D.~Neilsen, J.~Read, C.~Reisswig, M.~Ruffert (who acted
as referee), L.~Santamaria, E.~Schnetter, A.~Tonita, and S.~Yoshida is
also acknowledged. The computations were performed at the AEI and at
LONI.  This work is also supported by the DFG SFB/Transregio~7 and by
the JSPS grant 19-07803.


\begin{thebibliography}{50}

\bibitem[\protect\citeauthoryear{Abdolrahimi}{2009}]{Abdolrahimi2009}Abdolrahimi S.,
  arXiv:0905.0229 (2009)

\bibitem[\protect\citeauthoryear{Anderson et al.}{2008}]{Anderson2008} Anderson M.,
  Hirschmann  E.~W., Lehner L., Liebling S.~L., Motl P.~M., Neilsen D.,
  Palenzuela C. and Tohline J.~E., Phys. Rev. Lett.  100, 191101
  (2008)

\bibitem[\protect\citeauthoryear{Ant\'on et al.}{2006}]{Anton05} Ant\'on L., Zanotti O.,
  Miralles J.~A., Mart\'\i~J.~M., Ib\'a\~nez J.~M., Font J.~A. and
  Pons J.~A., Astrophys. J.  637, 296 (2006)

\bibitem[\protect\citeauthoryear{Baiotti et al.}{2008}]{Baiotti08} Baiotti L., Giacomazzo B.
  and Rezzolla L., Phys. Rev. D 78, 084033 (2008)

\bibitem[\protect\citeauthoryear{Colella and Woodward}{1984}]{Colella84} Colella P. and Woodward P.~R., 
  J. Comput. Phys. 54, 174 (1984)

\bibitem[\protect\citeauthoryear{Giacomazzo \& Rezzolla}{2006}]{Giacomazzo:2005jy}
  Giacomazzo B. and Rezzolla L., Journal of Fluid Mechanics 562, 223
  (2006)

\bibitem[\protect\citeauthoryear{Giacomazzo \& Rezzolla}{2007}]{Giacomazzo:2007ti}
  Giacomazzo B. and Rezzolla L., Class. Quantum Grav.  24, S235
  (2007)

\bibitem[\protect\citeauthoryear{Harten et al.}{1983}]{Harten83} Harten A., Lax P.~D. and
  van Leer B., SIAM Rev. 25, 35 (1983)

\bibitem[\protect\citeauthoryear{Koppitz et al.}{2007}]{Koppitz-etal-2007aa} Koppitz M.,
  Pollney D., Reisswig C., Rezzolla L., Thornburg J., Diener P. and
  Schnetter E., Phys. Rev. Lett. 99, 041102 (2007)

\bibitem[\protect\citeauthoryear{Lindblom et al.}{2008}]{Lindblom08} Lindblom L., Owen B.~J. and
Brown D.~A., Phys. Rev. D 78, 124020 (2008)

\bibitem[\protect\citeauthoryear{Liu et al.}{2008}]{Etienne08} Liu Y.~T.,
  Shapiro S.~L., Etienne Z.~B. and Taniguchi K.,
  Phys. Rev. D 78, 024012 (2008)

\bibitem[\protect\citeauthoryear{Meszaros}{2006}]{Meszaros:2006rc} Meszaros P.,
  Rept. Prog. Phys. 69, 2259 (2006)

\bibitem[\protect\citeauthoryear{Oechslin and Janka}{2007}]{Oechslin07b} Oechslin R. and
  Janka H.~T., Phys.Rev.Lett. 99, 121102 (2007)

\bibitem[\protect\citeauthoryear{Piran}{2004}]{Piran:2004ba} Piran T., Rev.
  Mod. Phys. 76, 1143 (2004)

\bibitem[\protect\citeauthoryear{Pollney et al.}{2007}]{Pollney:2007ss} Pollney D.,
  Reisswig C., Rezzolla L., Szil\'agyi B., Ansorg M., Deris B.,
  Diener P., Dorband E.~N., Koppitz M., Nagar A., et~al., Phys. Rev.
  D76, 124002 (2007)

\bibitem[\protect\citeauthoryear{Price and Rosswog}{2006}]{Price06} Price R.~H. and
  Rosswog S., Science 312, 719 (2006)

\bibitem[\protect\citeauthoryear{Read et al.}{2009}]{Read2009} Read J.~S., Markakis C.,
  Shibata M., Uryu K., Creighton J.~D.~E. and Friedman
  J.~L., Phys. Rev. D 79, 124033 (2009)

\bibitem[\protect\citeauthoryear{Schnetter et al.}{2004}]{Schnetter-etal-03b} Schnetter E.,
  Hawley S.~H. and Hawke I., Class. Quantum Grav.  21, 1465 (2004)

\bibitem[\protect\citeauthoryear{Taniguchi \& Gourgoulhon}{2002}]{Taniguchi02b} Taniguchi K.
  and Gourgoulhon E., Phys. Rev. D 66, 104019 (2002)

\bibitem[\protect\citeauthoryear{Toth}{2000}]{Toth2000} Toth G., J.  Comput. Phys.  161, 605
  (2000)

\bibitem[\protect\citeauthoryear{Urpin et al.}{1998}]{UGK98} Urpin, V. and Geppert, U. and
  Konenkov, D., MNRAS, 295, 907 (1998)


\bibitem[\protect\citeauthoryear{Yamamoto et al.}{2008}]{Yamamoto2008} Yamamoto T., Shibata M.
  and Taniguchi K., Phys. Rev. D 78, 064054 (2008)
\end{thebibliography}
\end{document}